\documentclass[global,twocolumn]{svjour}

\usepackage{graphics}
\usepackage[agsm]{harvard}

\journalname{Biological Cybernetics}

\begin{document}

\makeatletter
\renewcommand{\HAR@hisep@agsm}{;}
\makeatother

\title{Time delay and partial coherence analyses to identify cortical
connectivities}

\subtitle{}

\author{R. B. Govindan \inst{1}, Jan Raethjen \inst{1}, Kathrin Arning
  \inst{1}, Florian Kopper \inst{1} \and G\"unther Deuschl \inst{1} 
\thanks{\emph{Correspondence:} Prof. Dr. G. Deuschl \newline 
(e-mail:g.deuschl@neurologie.uni-kiel.de) \newline Department of Neurology, Christian-Albrechts
University of Kiel, Schittenhelmstrasse 10, D-24105 Kiel, Germany \newline FAX: 0049-431-5978502}} 

%

%

\institute{Department of Neurology, Christian-Albrechts
University of Kiel, Schittenhelmstrasse 10, D-24105
Kiel, Germany }

\date{Received: \today / Revised version: \today}


%

\maketitle

\begin{abstract}

Recently it has been demonstrated by Albo that partial coherence analysis is
sensitive 
to signal to noise ratio (SNR) and that it will always identify the signal with the
highest SNR among the three signals as the main (driving) influence. 
We propose to use time delay analysis in parallel to partial coherence
analysis to
identify the connectivities between the multivariate time series. Both are
applied 
to a theoretical model (used by
Albo) to analyse the connections introduced in the model. Time
delay analysis identifies the
connections correctly. We also apply these analyses to the electroencephalogram (EEG) and
electromyogram (EMG) of essential tremor patients and EEG of normal subjects while
bimanually tapping their index fingers. Biologically plausible
cortico-muscular and cortico-cortical connections are identified by
these methods.  
\end{abstract}

\section{Introduction}

\label{intro}

Coherence measures the degree of linear association (in frequency domain)
between the two signals \cite{amj1,hal1}.
The common influence of the third signal on the correlation between two
signals is addressed 
by partial coherence analysis
\cite{lopes,turbes,hal1,sherman,kocsis,mima1,timmermann,kubota,raethjen1}.
Understanding the role of more than one signal in the 
correlation between two signals has led to the graphical model
\cite{dahl1,ros1}. If the partial coherence between two signals is zero (is
below the pre-defined confidence limit), it is 
usually concluded that the presumed signal whose contributions are removed by
partial coherence, is responsible for the correlations (coherence) between
the two signals. Alternatively, the conclusion that the presumed signal has
\emph{caused} the other two signals has been drawn \cite{ger}. This idea has
been used by 
Gersch to 
identify the source of the epileptic foci \cite{ger}. But the concept of
attributing the causal (driver) nature to a signal when it abolishes the
coherence between the two signals upon partial coherence analysis, is
questioned in the recent past 
\cite{albo}. It has been demonstrated by \cite{albo} using an AR2 model as
source signal and two of its time delayed versions as response signals that partial
coherence analysis can yield spurious results. In this model, the source
signal namely the AR2 process is contaminated with the highest amount of
noise. In one of the time delayed versions of AR2, a slightly less amount of
noise is added. Yet another time delayed version of AR2 is left out as noise
free. This is labeled as Case I in \cite{albo}. Upon partial coherence analysis, they found that the signal with the highest
signal to noise ratio (SNR) which, in their case, is the last version stated above,
was incorrectly identified as driver signal. They provide further evidences for
skepticism of partial coherence technique by analysing the 
neuronal data (theta rhythms) recorded from different centers of the limbic
system of the rat. At the end they propose to use Granger causality
analysis \cite{berna,liang,baccala,kaminski} to identify the causal relation
between 
the different neuronal signals. 

From the above discussion it is clear that a zero value of partial coherence
between three signals need not necessarily imply that the connection between
the two signals are established only through the third signal. Even if the
results 
of partial coherence are correct, 
from a zero value of partial coherence, one cannot
understand the
nature (direction) of interaction between the signals. Time delay between the
signals 
allows to  
identify the direction of the information flows. By analysing the time delay
between the signals, we can verify the results of the partial coherence and
also we can understand the nature of the interaction between the signals.   
Though from the delay analysis we can understand the nature of
information flow, it will be difficult to judge whether or not the
connection between two centers is direct or indirect.  In the
foregoing discussion we ascribe the direction of information flow
between centers based on delay analysis and/or partial coherence
analysis. In an ideal situation both analyses should yield the same
results. Since the results of the partial coherence analysis are
susceptible to SNR of the signals, in an experimental setting, it will
be very difficult to rely on the results of one of the analyses
alone. In this study we use delay analysis in parallel to the partial
coherence analysis.     
In
this study we only consider the correlations between three signals. In
parallel we estimate the delay between the 
three distinguishable pairs. Based on the results of
these analyses, we establish the connection and direction (from delay
analysis) between the
three signals considered. Many physiological
questions are related only to processes happening in a particular frequency band and hence the natural selection will be frequency
specific methods. We use the maximising coherence method to identify the delay
between the two signals at a given frequency
\cite{carter,govindan,raethjen2}. Due to the time delay between the two
signals there will be (time) misalignment between them. Because of
this misalignment the coherence estimated between the two signals will be
slightly reduced. In order to compensate for this reduction in the coherence
and 
thereby to estimate the time delay, we artificially shift the time series by a
time lag. Coherence in a selected frequency band is estimated as a
function of lag. Coherence will increase as a function of lag and
reach a maximum value at the time lag equal to the time delay between the two
signals. 

First, we perform time delay and partial coherence analyses for the model used
in \cite{albo} to 
identify the connections (used) between the different variables of the
model. We find that the time delay analysis unambiguously identifies the
direction of interaction between different variables of the 
model.  Using these
techniques we analyse the electroencephalogram (EEG) and 
electromyogram (EMG) of essential tremor (ET) patients, to understand the direction
of 
interaction between the two tremor related cortical areas in the region of the
mesial premotor cortex close to the SMA and the region of the primary
sensorimotor cortex
(MC). In this case, partial
coherence analysis identifies that the connection between the two cortical
signals is
mainly established through the third (muscle) signal with the highest SNR
whereas the time delay analysis identifies entirely different connections. In
another application, we apply these techniques to the EEG recordings from  the
left and right sensorimotor  
cortical regions and midline region of normal subjects
during bimanual finger movements to understand the flow of the movement related
information. In one of these examples (all cortical areas show comparable
SNR), both analyses yield conforming results while in the other example again
the results of delay analysis differ from the partial coherence analysis. 
\section{Materials and Methods}

\label{sec2}

\subsection{Data acquisition}

\label{sec3}

Details of spectral analysis of ET can be found in 
\cite{raethjen2}.
This study was approved by the local ethics committee and all the subjects gave
informed consent. 
In a dimly lit room, subjects are asked to sit on a comfortable chair in a
slightly supine position with both hands held against gravity (for normal
subjects the hands are kept in a comfortable position to execute rhythmic
bimanual movement of
the index fingers) while the
forearms are supported. EEG is recorded with 64-channel 
EEG system (Neuroscan) with standard electrode positions \cite{klem}. Surface
EMG is recorded bipolarly from EMG electrodes attached to wrist extensors and flexors of
both arms. In the case of normal subjects during the bimanual finger tapping
the EMG electrodes are positioned on the index extensors and flexors muscles
of the forearm. EEG and EMG data are sampled
at a rate of 1000 Hz. EEG and EMG are band pass filtered on line between 0.01 to 200 Hz and 30-200 Hz, respectively. Data are stored in a computer and are
analysed offline. Each recording lasted for 1-4 minutes. Artefacts like eye
blinks, base-line shift, etc., are discarded by visual inspection. Further,
EMG signals are full wave rectified (magnitude of the deviations from mean
value) and EEG is made reference free by constructing second (spatial)
derivative, Laplacian \cite{hj1,hj2}.
\subsection{Methods}

\label{sec4}

Let $x(t)$, $y(t)$ and $z(t)$ be the three simultaneously measured signals of
length $N$.  Let the 
sampling frequency be $s$ Hz. Following \cite{hal1}, we 
calculate coherence and partial coherence as follows: We divide the
signals in to $M$ disjoint time segments of length $L$, such that
$N=LM$. We calculate, power spectra $S_{xx}$, $S_{yy}$, $S_{zz}$ and
cross-spectra $S_{xy}$, $S_{xz}$ and $S_{yz}$
in each of the disjoint windows. Finally, we average these quantities 
across all the segments to get the estimate of the same.
We estimate coherence between the signals $x$ and $y$ as follows:
$
\widehat{C_{xy}(\omega)}~=~{\left|\widehat{S_{xy}(\omega)}\right|^2}/{\widehat{S_{xx}(\omega)}
\widehat{S_{yy}(\omega)}}~.$
Similarly, we estimate the partial coherence between the three signals $x(t),
y(t)$ and $z(t)$ as follows:
$$
\widehat{C_{xy|z}(\omega)}~=~\frac{\left|\widehat{CY_{xy}(\omega)}-\widehat{CY_{xz}(\omega)}\widehat{CY_{zy}(\omega)}\right|^2}{(1-\widehat{C_{xz}(\omega)})(1-\widehat{C_{zy}(\omega)})},
$$ where $\widehat{CY_{ij}(\omega)}$ is a complex valued function whose
magnitude is called coherency \cite{hal1} between the two signals $i$ and
$j$. The over-cap indicates the estimate of that quantity. In the following
discussions, for the sake of convenience, we use these quantities without
over-cap. The confidence limit for coherence at 100\% $\alpha$ is
$1-(1-\alpha)^{1/(M-1)}$. Similarly the confidence limit for partial coherence
at 
100\% $\alpha$  is $1-(1-\alpha)^{1/(M-2)}$ \cite{brillinger,hal1}. In our
calculations we have set $\alpha=.99$. Any value of coherence above the
confidence limit is considered to show significant correlation between the
time series and 
the value of coherence below this confidence limit indicates lack of
correlation between the two time series. For the interpretation of partial
coherence we refer to Sect.~\ref{intro}. In the above spectral
estimations, we choose a segment length of $L=1s$ which results in a frequency
resolution of 1 Hz for the estimated spectral quantities. However, one can
choose different segment lengths to have a good compromise between sensitivity
and reliability of the estimated quantities.

We extend the coherence analysis to estimate the time delay between the two
signals $x(t)$ and $y(t)$. Analogous to
cross-correlation, if there is a time delay $\delta$ between the two signals,
the value of coherence will be slightly reduced. In order to compensate for
this 
reduction in the coherence due to $\delta$, we shift one of the time series
say $x(t)$ (assuming $x(t)$ is the time delayed version of $y(t)$), by a 
constant time lag $\tau$. We discard the extra $\tau$ points (in sample units)
in $y(t)$ to have the same length for both time series. We estimate the
coherence  $C(\tau)_{\omega_0}$ in a 
selected frequency band $\omega_0$ as a function of $\tau$. The value of
$C(\tau)_{\omega_0}$ will increase as a function of $\tau$ and reach a maximum
value when $\tau=\delta$. This procedure is repeated for the other time series
$y(t)$ to estimate the time delay from $x(t)$ to $y(t)$. This idea of
estimating the time delay by maximising the coherence has been successfully applied to
acoustic signals \cite{carter}. A similar idea using phase synchronisation,
has been
employed to estimate the time delay between and among different
meteorological variables \cite{diego}. Since coherence measure is
sensitive to the length of the data, we discard the points corresponding to
maximum $\tau$ from both time series, and consider only the length of the time
series which is integer multiple of the segment length $L$. In this way
$C(\tau)_{\omega_0}$ estimated at different values of $\tau$ will have the same
confidence limit. This ensures that the  
maximum value of the coherence is reached only because of the time delay and
not because of the different lengths of the data used in the estimation of the
$C(\tau)_{\omega_0}$. 

We have used the confidence limit (see above) to assess the significance of
$C(\tau)_{\omega_0}$. For the theoretical model which is AR2 and its time
delayed variants, one can 
obtain the error-bar of the delay from standard deviations of the delays
estimated for several realisations of the AR2 process. In order
to get the error-bars of the estimated delay 
(variability of the delay) for the biological data, we use surrogate analysis
\cite{kantz}. We 
synthesise surrogate data for this purpose by exploiting one of the basic
assumptions 
of spectral analysis: that the data in disjoint segments are independent
\cite{hal1}. We shuffle the disjoint segments of the time series from which
the original spectrum is estimated. This is done only for one of the time
series from 
which  
time delayed information is assumed to flow to another time series. Then, we
repeat this procedure for the other time series. Thus, in
this type of surrogate the 
original spectrum of both the time series is preserved but only the cross
spectrum will be different. We prepare 19 different surrogates for each of our
analyses and calculate the time delayed coherence function
$C(\tau)_{\omega_0}^{sur}$ for each of the realisations. We make a null
hypothesis that $C(\tau)_{\omega_0}$ calculated for
the original time series $x(t)$ and $y(t)$ is due to spurious correlations. We
estimate the significance of the difference $S(\tau)$ between
$C(\tau)_{\omega_0}$ 
and $C(\tau)_{\omega_0}^{sur}$ as $S(\tau)=\left|C(\tau)_{\omega_0}-<C(\tau)_{\omega_0}^{surr}>
\right|/\sigma[C(\tau)_{\omega_0}^{surr}]$, where $<.>$ indicates the average
over all the surrogates and $\sigma[.]$ represents the standard deviation
between 
different realisations. If $S(\tau)>2$ then we reject the null hypothesis that
$C(\tau)_{\omega_0}$ is due to spurious correlation
\cite{kantz,theiler}. Though we compute $S(\tau)$ for all values of $\tau$ at which
$C(\tau)_{\omega_0}^{sur}$ and  $C(\tau)_{\omega_0}$ are estimated, for
simplicity we report
here only $S(\tau=\delta)$. When
the null hypothesis is rejected, we calculate the error-bars of the delay as
follows: We subtract  $C(\tau)_{\omega_0}^{sur}$ from $C(\tau)_{\omega_0}$ and
estimate delay for each surrogate subtracted realisation. We report the mean
value 
of these delays as the delay
between the two time series and their standard deviation as the
error-bar of the delay. For the sake of clarity, we plot 
\begin{eqnarray}
C'(\tau)_{\omega_0}&=&\left[C(\tau)_{\omega_0}-<C(\tau)_{\omega_0}^{surr}>\right]
\nonumber \\
& & -\left[C(\omega_0)-<C(\omega_0)^{surr}>\right] \nonumber.
\end{eqnarray} 
Note that $C(\omega_0)$ is
the value of the coherence at
$\tau=0$ and $<C(\omega_0)^{surr}>$ is the average of all the surrogates
at $\tau=0$. By this definition, $C'(\tau)_{\omega_0}$ will pass through a
zero value at $\tau=0$ and reach a maximum value at $\tau=\delta$. Based on
the above arguments we calculate the delay between the two time series as
$$\delta={\max_\tau}~C'(\tau)_{\omega_0}.$$ The preference of this method over
other conventional methods of delay estimation is discussed in
\cite{govindan}. It has been shown by Carter \cite{carter} that the
uncertainty in the estimation of coherence due to the time delay between the
two time series is proportional to $1/L$ (where $L$ is the \emph{FFT length} used
in the estimation of the spectral quantities). In order to favor the
estimation of the time delay, we increase the uncertainty in the coherence
estimation by decreasing $L$
(which in turn results in the poorer frequency resolution of the spectral
estimates). Sometimes, decreasing the frequency resolution may result
in the loss of coherence (as the spectral quantities are poorly estimated)
which should be checked before proceeding to the delay analysis. In the
forgoing 
analysis, we have used a frequency resolution of 5 
Hz for the theoretical model and a frequency resolution of 2 Hz for the
biological systems.  

As this method relies on the bias (due to delay) in the (coherence) estimator, results obtained might be sensitive to the other factors causing
the bias in the coherence estimate. In addition to delay, the other two dominant factors which can affect coherence estimator are the FFT length $L$ (see above) and the 
SNR of the signals. As discussed above the \emph{FFT length} is coupled to the
delay \cite{carter} in causing the bias in the coherence estimate. By properly 
choosing this value (usually in the range of 1/5 sec to 1/2 sec) one can
minimise the effect due to this bias. The bias caused by SNR cannot be
addressed directly. However, the error-bars of the delay estimate to some
extent can reflect the effect of SNR on the delay estimate. Further, length of
the dataset can cause bias in the coherence. For a given system, the number of
points for the time shifted coherence is kept the same (for all pair
combinations, see above) to avoid this bias.  
Since all the time shifted coherence analysis is done for the same number of points (see above) it will not affect the final results. 

\section{Application to Model system}

\label{sec5}

In this section we test our hypothesis that delay can serve as a measure to
identify the direction of interaction, in a group of three
coherent signals and can overcome the methodological pitfalls inherent to
partial coherence analysis \cite{albo}. We choose to use the model used in  \cite{albo} which is an AR2
process (source) and its time delayed versions as response signals. AR2
processes have been used to model biological processes
\cite{honerkamp,timmer1,timmer2}. The model is given by:
\begin{eqnarray}
X_1(t)&=&X_0(t-\delta_1)+\eta_1(t) \nonumber \\
X_2(t)&=&X_0(t)+\eta_2(t) \nonumber \\
X_3(t)&=&X_0(t-\delta_2)+\eta_3(t), \nonumber 
\end{eqnarray}
 where $X_0(t)=0.8X_0(t-1)-0.5X_0(t-2)+\eta_0(t)$ is a AR2
 process. $\eta_i(t)$ are Gaussian white noise processes with zero mean and
 unit variance. We assume 1000 Hz as sampling frequency. We set
$\delta_1=3$ ms, $\delta_2=5$ ms, $var(\eta_1(t))=0.04$, $var(\eta_2(t))=0.06$
 and 
$var(\eta_3(t))=0.0$. In this model, there is a delayed flow of information
 from 
 $X_2(t)$ to $X_1(t)$ and $X_3(t)$ with a delay of 3 and 5 ms,
 respectively. Also there is delayed information flow from $X_1(t)$ to
 $X_3(t)$ with a delay of 2 ms.
By construction of the model, $X_2(t)$ is the mediator as it
is solely responsible for the existence of (establishing the connection
between) $X_1(t)$ and $X_3(t)$.   

 Results of the delay, coherence and partial coherence 
 analyses for this model are given in Fig. 1. Figures 1(b,c,f)
 show a significant coherence between the three time series. 
 Figures 1(d,g,h) display partial coherence between the time series
 $X$ and $Y$ accounting for the common influence of the
 third time series $Z$ and is indicated by $X-Y/Z$.   



{\bf{Figure 1 will appear here}}.

In Fig. 1d the partial coherence between $X_1(t)$ and $X_2(t)$ accounting for
the common influence of $X_3(t)$ is insignificant seemingly indicating that the
connection 
between $X_1(t)$ and $X_2(t)$ is mainly established through $X_3(t)$. In
Fig. 1(g,h) there is a decrease in the (partial) coherence (compared to their
coherence 
shown in Fig. 1(c,f), respectively) when the influence of the
third signal is removed. One may conclude that the third signal in Fig. 1(g,h)
also shares some of its signal with the other two variables but is not solely
responsible for the correlation between these two. Thus, the partial coherence
analysis already indicates 
that all signals are inter connected but $X_3(t)$ with the highest SNR is
identified as the main influence (mediator) to establish the connection
between the other variables within this network. The direction of interaction
introduced in the model 
is shown in Fig. 1j.  The
connections identified by partial coherence analysis are shown 
in Fig. 1k. 
 
Now, we consider
the results of the delay analysis shown in Fig. 1(a,e,i). 
In Fig. 1a there is a delay of $3\pm0.35$ ms from
$X_2(t)$ to $X_1(t)$ and the significance of deviation from the surrogates is
$S(\delta)=69.3$. In Fig. 1e there is a delay of $2\pm0.28$ ms from $X_1(t)$
to $X_3(t)$ and the significance of deviation from the surrogates is
$S(\delta)=102.87$. In Fig. 1i there is delay of $5\pm0.47$ ms from $X_2(t)$ to
$X_3(t)$ and the significance of deviation from the surrogates is
$S(\delta)=91.24$. In all these cases, $S(\delta)$ is greater
than 2 standard deviations indicating that the correlations in $C(\tau)_{\omega_0}$ are not
spurious. From delay analysis, it is clear that there is flow of information
from $X_2(t)$ to $X_1(t)$ and $X_3(t)$ with a delay of 3
and 5 ms, respectively. The fact that $X_1(t)$ and $X_2(t)$ both project to
$X_3(t)$ but no information flows from $X_3(t)$ to $X_1(t)$ or $X_2(t)$
clearly rules out that $X_3(t)$ is the main influence as suggested by the
partial coherence analysis. Conclusion based on the delay
analysis is shown in Fig. 1l.  

We also have repeated the analysis for slightly
higher 
intensity of noise to check the robustness of the delay
analysis to 
noise.  There is no change in the final conclusion drawn in Fig. 1l except
for the 
increase in the error-bar of the delay estimate.  The results of the
partial coherence analysis remain consistent with
Fig. 1(k).
 Based on the delay values, in Fig. 1k, we draw the arrows from
  source (which is X2 signal) to targets (which are X1 and
  X2). Further there is a delayed flow of information from X1 to X3,
  and hence we have drawn an arrow from X1 to X3.  Although partial
  coherence analysis has wrongly identified X3 as the mediator between
  the two signals (Fig. 1d), a slight reduction in the (partial)
  coherence values in Fig 1 g and h when X3 is partialised for X1 and
  X2 (compare to Fig. 1c and 1f, respectively) indicates that X1 and
  X2 are also directly connected or share a common source with X3.  If
  X3 had been the sole source as identified by the partial coherence
  analysis in Fig. 1d, its coherence with the other two signals
  should not have diminished when partialised for the third signal. Thus,
  based on the delay analysis and from the above discussion we
  establish the connectivities between the variables of the model as
  shown in Fig. f.

If we consider an another different situation in the above model, with $var(\eta_2=0)$ and $var(\eta_1),var(\eta_3)$ are
  non-zeros, and with $var(\eta_3)>var(\eta_1)$, then the results of the partial
  coherence will clearly identify that the connection between X1 and 
  X3 are established through X2 indicating that X2 is the source
  signal.  Delay analysis will also yield similar results. However, in 
  this case, as well as in the above scenario, though we can argue the
  connection (direction of the information flow) between two centers 
  based on the delay analysis we cannot clearly judge whether or not 
  they are directly or indirectly coupled. Thus, the results of both 
  analyses should be interpreted with great care in the presence of 
  cross-channel relations.   
 
 Further, for the above model, 
one can obtain 
similar results by cross-correlation analysis or by phase synchronisation
analysis \cite{diego}. But for biological data, we
are interested in the nature of interaction 
between the two signals in a particular frequency band. So in the
next section we continue to use the maximising coherence method. 
\section{Application to Biological systems}

\label{sec6}

In this section we perform the delay and
partial coherence analyses to identify the connection between the cortical and
muscle signals by applying them to (i) the
EEG and EMG time series in ET patients (ii) 
the EEG of normal subjects while performing rhythmic bimanual tapping of their
index 
fingers. We also show that the results of 
partial coherence analysis tend to identify the signal with the highest SNR as
the mediator through which the connection between the other two signals is
established.  We calculate the signal to noise ratio (SNR) of the signal
by considering the ratio of the power at the desired frequency to noise
level. We use the mean value of power in the frequency 
range of 100 to 200 Hz as noise
level (upper cut off of our filter is at 200 Hz).  Connections and nature of
interactions identified by delay 
analysis are different from those identified by partial coherence analysis and
provide biologically conceivable interpretation.  

Before proceeding any further, we construct cortico-muscular isocoherence maps
\cite{raethjen2,govindan} for all the cases discussed below. In the case of
ET, there are two distinct regions (hot spot) with significant cortico-muscular coherence
at the tremor frequency, one in the frontal and midline regions, and another
in the region of the contralateral primary sensorimotor cortex. In the normal
subjects the  
voluntary rhythms of both hands are represented in the right and left
sensorimotor areas and also in the frontal/midline region. As in the tremor
patients this secondary hot spot likely reflects an involvement of premotor
cortical areas (e.g.) SMA \cite{raethjen2}. In both cases we
take an electrode from each region displaying maximum cortico-muscular
coherence to identify the connections between the different cortical areas. 
   
\subsection{Application to ET}

\label{sec7}

ET is a common movement
disorder characterised by postural tremor of the arms \cite{deuschl}. Other
neurological abnormalities are typically absent in essential tremor
\cite{findley}. Experimental studies on animals \cite{llinas,lamarre} and on
human beings \cite{jenkins,hallett,bucher} show 
that different parts of the brain are involved in essential tremor. Evidence
for the involvement of the thalamus in the tremor oscillations is
also known \cite{hua,benabid,schuurman}. The role of cortical motor centers is
currently under debate \cite{hal2,raethjen2,hellwig1}. We 
consider two ET patients which we label ET1 and ET2. These two subjects
display cortico-muscular 
coherence at the tremor frequency (4 and 5 Hz) in a frontal area close to the
midline, possibly supplementary 
motor area (SMA)  
and in the area of the sensorimotor cortex (MC). We are interested in the
tremor related connectivities of these two motor regions with each other and
with the periphery. For ET1, results of delay, 
coherence and partial coherence analyses are given in Fig. 2.  

{\bf{Figure 2 will appear here}}.

The muscle spectrum in Fig. 2j shows tremor
oscillations around 4 Hz. Fig. 2b displays the cortico-cortical coherence between F2 (SMA) and C2
(MC). There is a significant coherence at the tremor frequency of 4
Hz. Fig. 2(c,f) display a significant cortico-muscular coherence at the tremor
frequency of 4 Hz \cite{mima2,mima3,mima4,hal2}. In Fig. 2d, (partial)
coherence between F2 and C2 accounting for the common influence of EMG is
vanished at the tremor frequency of 4 Hz seemingly indicating that the external tremor
is mainly responsible for the coupling between these areas. However, in
Fig. 2(g,h) partial 
cortico-muscular coherence accounting for the other cortical area is
also slightly reduced compared to the coherences shown in Fig. 2(c,f).
This indicates that these two regions are also coupled
\cite{raethjen2}. Thus the partial coherence analysis seems to support the
notion that all recording sites are connected in a motor network with the
peripheral tremor being the leading influence of tremor related activity. SNR
of F2, C2 and EMG are 1.52, 
10.25  
and 70.58, respectively. By comparing the SNR values we see that partial coherence analysis
has identified the signal with the highest SNR as the main mediator within the
cortico-muscular network as predicted by the model \cite{albo}. 

To establish the direction of the different connections we again perform the
delay analysis.
We choose the tremor frequency of 4 Hz as $\omega_0$. Fig. 2a shows a delay of
$7\pm0.67$ ms from C2 to F2 and $3\pm0.88$ 
ms from F2 to C2 and the significances of deviation from the
surrogates are $S(7)=4.72$ and $S(3)=3.36$.  
Fig. 2e shows a
delay of $19\pm9.78$ ms from F2 to EMG and the significance of deviation from
the surrogates is $S(19)=2.86$. 
Fig. 2i shows a delay of $9\pm3.83$ ms from EMG to C2 
and $11\pm3.71$ ms from C2 to EMG and the significances of deviation from the
surrogates are $S(9)=15.82$ and $S(11)=15.83$. Significance of deviation from
the surrogates in all the cases are greater than 2 standard deviations indicating
$C(\tau)_{\omega_o}$ is not due to spurious correlations. There is a
uni-directional flow from
F2 to EMG and a bi-directional flow between F2 and C2. Also there is a
bi-directional flow between C2 and EMG. Thus the cortical areas both influence
each other while the EMG only exerts some influence on one of the cortical
sites. This dominance of cortico-muscular flow and relative lack of
musculo-cortical interaction indicates that the EMG cannot be the main
influence responsible for the cortico-cortical interaction as inferred from the
partial coherence analysis. On the contrary the two cortical areas mutually
exchange information and both project to the EMG thereby contributing to the
peripheral tremor rhythm. Cortico-cortical connections and the
cortico-muscular 
connections based on the delay analysis are shown in Fig. 2l. For the details
of 
discussion of cortico-muscular and  
musculo-cortical delays we also refer to \cite{govindan}. 

{\bf{Figure 3 will appear here}}

The results of delay, coherence and partial coherence analyses for ET2 are
shown in Fig. 3. As seen for ET1, there is a significant cortico-cortical
coherence between FCZ (SMA) and C4 (MC) (see Fig. 3b) at the tremor frequency of 5 Hz shown
in Fig. 3j. Also, there is a significant cortico-muscular coherence for both
SMA (FCZ) and MC (C4) as shown in Fig. 3(c,f) at the tremor frequency of 5
Hz. In Fig. 3d 
partial coherence between FCZ and C4 accounting for the common influence of EMG
is insignificant again seemingly indicating that EMG is mainly responsible for the
tremor related correlations between FCZ and C4. As seen in Fig. 2(g,h) for
ET1, also for ET2 (see Fig. 3(g,h)), the partial cortico-muscular coherence
accounting for the common influence of
the other cortical region is significant but has slightly
reduced compared to their corresponding coherences shown in Fig. 3(c,f). This
again shows (as concluded for ET1) that these two regions also seem to couple
leading to a reduction in the coherence when the influence of
the other  
cortical region is removed. The SNR values for FCZ, C4 and EMG are
20.25, 26.64 and 161.22, respectively. Thus the partial coherence has again
identified the EMG signal with the highest SNR as the mediator.      

For delay analysis, we choose $\omega_0$ as 5 Hz which is the tremor frequency.
In Fig. 3a, there is a delay of $3\pm1.92$ ms from FCZ to C4 and the
significance of deviation from the surrogates is $S(3)=14.06$. In Fig. 3e, there is a delay of $25\pm10.42$ ms from FCZ to
EMG and the significance of deviations from the surrogates is $S(25)=5.05$. In Fig. 3i, there is a
delay of $18\pm9.21$ ms from C4 to EMG and the significance of deviation from the
surrogates is $S(18)=13.01$. In all the cases the significance of deviations
from the surrogates is greater than 2 standard deviations indicating that the
correlations in $C(\tau)_{\omega_0}$ are not spurious. There is a clear
uni-directional flow from FCZ to 
C4 in Fig. 3a. Also there is uni-directional flow from FCZ to EMG in
Fig. 3e. In Fig. 3i there is uni-directional flow from C4 to EMG. There is no
flow from EMG to the cortical centers to account for the
cortico-cortical connection as suggested by the partial coherence
analysis. Based on the delay analysis, the cortico-cortical connections and
the cortico-muscular connections are given in Fig. 3l and this conclusion is
different from the conclusion drawn from partial coherence analysis (see
Fig. 3k). 

In both cases, there is a cortico-muscular delay of 11-20 ms from MC to
muscle. This is in keeping with the experimentally observed conduction time of
$15\pm2$ ms between cortex to muscle \cite{rothwell}. The longer
cortico-muscular delay of 19-25 ms between the frontal (SMA) region and
muscle may indicate a different pathway or  mode of interaction with the
periphery.  The musculo-cortical delay (EMG-MC) found in ET1 is in keeping with normal
latencies of somatosensory evoked cortical potentials and indicates that the
peripheral tremor is also fed back to cortex. The
cortico-cortical delay of 3-7 ms between SMA and MC is in agreement with the
delay of 3-6 ms between premotor stimulation and its effect on the motor
cortex recently studied with transcranial magnetic stimulation \cite{civardi}.

In the above examples, EMG has the highest SNR and hence it is
identified as the main mediator, through which the (tremor related) cortical connections are
established, by partial coherence analysis. This would imply feedback of the tremor rhythm to the
cortex as the main mechanism for the cortico-muscular and cortico-cortical
coherence.  But, based 
on the delay analysis we can clearly show that the SMA and MC
both share 
tremor related activity themselves which in turn is transmitted
to the muscle.  This is an argument that the cortex being part of the central
generating network of ET \cite{hellwig1,raethjen1}. While the disappearance of
the cortico-cortical 
coherence when the common influence of the EMG is removed, is a methodological artefact the
remaining albeit slightly reduced cortico-muscular coherence after
accounting for the influence of the other cortical signal is interpretable. It
indicates a 
somewhat independent connection of both areas with the peripheral tremor
possibly 
intermittently sharing their tremor related activity \cite{raethjen2}. This tremor related cortico-cortical interaction is bidirectional in
the first example. In the second presented case the main information flow
seems to be from SMA to MC. In summary, this study provides evidence that tremor related
correlations are transmitted from SMA or MC to muscle, 
but due to the low (tremor related) signal content in the cortex this
connection is not correctly identified by partial coherence analysis. 
  
\subsection{Application to cortical activity related to bimanual rhythmic
movements in Normal subjects}

\label{sec8}
 
In this section we perform the time delay analysis to identify the interaction
between the left and right sensorimotor cortices and the midline
area of healthy subjects during bimanual rhythmic movements of the index fingers. This
study will throw light on the cortical network 
involved in bimanual movements. In a recent  study \cite{pollok}, it has
been shown that the rhythm of voluntary unilateral hand movements is
represented in the contralateral sensorimotor cortex as part of the central
generating network of these voluntary rhythms. For bimanual movements it has
been postulated that the cortical midline areas (especially SMA) play a
major role \cite{jancke,lang,stephan,donchin,immisch}. On the other hand the
interhemispheric connections via the corpus callosum between the primary
sensorimotor cortices on both sides seem to be important
\cite{eliassen,brinkman,andres}. As the more bilaterally organised SMA and the
more lateralized primary sensorimotor cortices are also tightly linked by
projection fibers, one may postulate that all three cortical areas take part in
the control of bilateral movements. It has recently been proposed on the basis
of monkey experiments that the SMA is not a superordinate center for bimanual
movement coordination but only part of an interconnected cortical network in
both hemispheres \cite{kazennikov}. 

We consider two subjects, hs1 and hs2. All the cortical areas (electrodes)
used for this study displayed significant cortico-muscular coherence at the
tapping frequency. But in this study we are interested in the movement related
cortico-cortical
connections. Again we compare the results of the
delay analysis with the partial coherence analysis. The results of partial
coherence analysis are also compared with the SNR of the cortical signals.

For hs1, the results of the delay, coherence and partial coherence analyses are
given in Fig. 4. There is a significant synchronised activity between the two
index finger extensor and flexor muscles around 4 Hz as displayed by their coherence spectrum in
Fig. 4j. Fig. 4(b,c,f) display significant cortico-cortical coherence at
the tapping frequency of 4 Hz (see Fig. 4j) as well as in other frequency
bands. In Fig. 4d, the partial coherence between the cortical signals from the
left and right hemispheres, accounting for the common influence of the midline
area (represented by CPZ) is insignificant at the tapping frequency as well as
in the other frequency bands. This seemingly indicates that the tapping
related activity in the
two hemispheres is coupled through the midline area (CPZ). In
Fig. 4(g,h) the partial coherences between one of the hemispheres and the
midline accounting for the influence of the other hemisphere remain significant but
have also slightly reduced compared to  coherence shown in
Fig. 4(c,f), respectively. This indicates that these areas also share some tapping related and
other activities between them. Based on these arguments one would conclude in
Fig. 4k that the midline region is the main mediator bringing in the coupling between the two
hemisphere. However, the reduction of the partial coherence accounting for the
influence of the lateral cortical areas (C4/C3) indicates that they also have
direct connections. The SNR values for C3, C4 and CPZ
are 51.28, 53.02 and 82.59, respectively. Thus, the partial coherence
analysis detects the recording site with the highest SNR as the leading
influence among the three
\cite{albo}. Now we look at the 
results of the delay estimation.   

{\bf{Figure 4 will appear here}}

We choose $\omega_0$ as 4 Hz at which there is a synchronised activity between
the two fingers. In Fig. 4a there is a delay of $6\pm0.88$ ms from C4 to C3
and the significance of deviation from the surrogates is $S(6)=78.64$. In Fig. 4e, there is a delay of
$2\pm0.41$ ms
from CPZ to C3 and the significance of deviation from the surrogates is
$S(2)=89.79$. In Fig. 4i
there is a delay of $3\pm0.74$ ms from C4 to CPZ and the significance of
deviation from the surrogates is $S(3)=150.2$. In all the cases, the
(significance of) deviation from surrogates is greater than 2 standard
deviations indicating that $C(\tau)_{\omega_0}$ obtained in all the cases are
not spurious. In Fig. 4a there is a uni-directional flow from C4 to
C3. Fig. 4e, shows a uni-directional flow from CPZ to C3. In Fig. 4i there
is a uni-directional flow from C4 to CPZ.  Based on these
results, we can conclude that 
there is a movement related information flow directly from right to left
hemisphere via
the midline area. Interestingly, the sum of the delay from C4 to CPZ and from
CPZ to C3 (5 ms) is very close to the direct delay between C4 and C3 (6
ms). Thus, we cannot exclude that the transmission between C4 and C3
calculated by maximising coherence only reflects the indirect transmission via
the midline area (CPZ) which was detected as the main mediator in the partial
coherence analysis. However, the fact that partial coherence results are also
in keeping with direct connections between C4 and C3 and that the midline has
the highest SNR are arguments against the midline being the sole factor in
connecting the three cortical areas. In fact, the delay analysis shows that
the midline area (CPZ) only projects to C3 while C4 projects both to C3 and
CPZ thus being the most influential area within this network of three cortical
areas. Connections
based on the delay estimation are shown in Fig. 4l.

Next we consider hs2 for which the results of the delay, coherence and partial
coherence analyses are given in Fig. 5. In an earlier work \cite{pollok}, it has been shown
that healthy subjects while performing bimanual movements exhibited
significant cortico-muscular coherence at the movement frequency and also at
double the movement frequency.  The coherence at both frequencies are
considered to be the movement related cortical representations. In this subject, we observe cortico-cortical coherence at the
tapping frequency and/or at double the tapping frequency. Based on the
results of earlier work \cite{pollok}, we interpret a complete loss of
coherence between two cortical areas at both the frequencies (tapping and
double the tapping frequency) when the common influence of the third cortical
area is removed, as the insignificant partial coherence. There is a significant coherent
rhythm in both  
index fingers at 3-6 Hz shown in
Fig. 5j.  In Fig. 5(b,c) there is a significant cortico-cortical coherence
between midline and lateral electrodes that showed coherence with the
peripheral voluntary movement rhythm. However the coherence is significant
between the hemispheres only at double the tapping frequency (6-7 Hz) (see
Fig. 5f).  In Fig. 5h, the partial coherence between the two hemispheres
accounting for the common influence of the midline region is insignificant,
seemingly indicating that the connection between the hemispheres are
established through 
the midline. In Fig. 5d there is a reduction in the partial coherence at
double the tapping frequency (compare to Fig. 5b), between FCZ and C3 when the
common influence of 
C4 is removed but the coherence at the tapping frequency remains almost the
same. In Fig. 5g, the partial coherence remains almost the same (see
Fig. 5c) at the 
basic as well as double the tapping frequency
between FCZ and C4 when the common influence of C3 is removed. The
SNR of FCZ, C3 and C4 are 44.07, 29.65 and 35.01, respectively. They 
are almost comparable to each other. Fig. 5k shows the main cortico-cortical
connections based on the results of partial coherence analysis.

{\bf{Figure 5 will appear here}}.

We choose $\omega_0$ as 6 Hz as there is significant coherence at this
frequency in all the three pairs. In Fig. 5a there is a delay of $6\pm1.83$ ms
from C3 to FCZ and a delay of $15\pm3.3$ ms from FCZ to C3. The significances
of deviation from the surrogates are $S(6)=9.86$ and
$S(15)=10.26$, respectively. In Fig. 5e, there is delay of $16\pm6.17$ ms from FCZ to C4 and
the significance of deviation from the surrogates is $S(16)=15.61$. In
Fig. 5i, there is a delay of $14\pm6.29$ ms from C4 to C3 and the significance
of deviation from the surrogates is $S(14)=5.68$. In all the cases, the
significance of deviation from the surrogates is greater than 2 standard
deviations indicating that the
correlations in $C(\tau)_{\omega_0}$ are not spurious. In Fig. 5a, there is a
bi-directional flow between FCZ and C3.  In Fig. 5e, there is a
uni-directional flow from FCZ to C4. In Fig. 5i, there is a uni-directional
flow from C4 to C3. Based on these arguments the cortico-cortical connections
with the direction of the information flows are given in Fig. 5l. Based on the results of partial coherence analysis, one
would expect a tapping related synchronisation between FCZ and C4 and that 
C3 gains access to C4 mainly through FCZ.  This is in keeping with the
connections 
obtained from the delay analysis. Again the reduction in the partial coherence
between FCZ and C3 accounting for the common influence of C4 indicates that
there also is a direct connection between C4 and C3 with a flow mainly from C4
to C3. This is exactly what we find in the delay analysis. Thus, in this case the results obtained from
both the analyses are not contradictory because the SNR at the three recording
sites are comparable and are in accordance with our starting hypothesis the
partial coherence analysis yielded valid results in this case.

Taking in to account the relatively large variability (error-bars), the delay
between the right and left sensorimotor cortices of 6 and 14 ms is in
keeping with the well known latencies of transcallosal inhibition around 8-15
ms \cite{meyer,gennaro1,gennaro2,schmierer}. The delay between the midline
area and the sensorimotor cortex is slightly shorter than the transcallosal
delay which is in keeping with the shorter latencies of inhibitory effects on
the primary motor cortex evoked by premotor cortex stimulation (3-6 ms) in
humans \cite{civardi}.

The results of our analyses in these two cases show that all three (bilateral
sensorimotor cortices and midline area) cortical areas are connected and that
the midline area (possibly SMA) is not necessarily the leading source of the
activity related to bimanual rhythmic movement. Thus our data support the
hypothesis that a bilateral interconnected cortical network including midline
structures (e.g. SMA) is involved in bimanual coordination \cite{kazennikov}. 
\section{Conclusion}

In order to understand the complex interactions between 3 or more signals, partial coherence is used as a potential tool. 
But partial coherence analysis is sensitive to SNR \cite{albo}. On the other hand, for data from multichannels, multivariate
analysis has been shown to perform well in terms of identifying the complex
interaction patterns underlying the system compared to the  
pair-wise analysis \cite{kus}.
In many physiological situation there will be delay between the signals
\cite{timmer,timmermann,govindan,pollok}. In this work, as a complementary
approach to partial coherence analysis, we use delay analysis to identify the
connections and the nature of interaction between the signals. 

In cases where results of partial coherence are affected by SNR, we
establish the connections between centers based on the delay. As
mentioned in the introduction, though we can argue about the direction of
information flows based on the delay analysis, it cannot distinguish 
between direct or indirect connection. However, as shown in our 
examples delay analysis can often help to detect spurious overall 
results of the partial coherence analysis. In these cases some 
reduction in partial coherence compared to coherence is typically seen 
also for those two connection that remained after partialization. This 
hints at a weaker direct connection also between those two centers 
for which partial coherence became insignificant, and it justifies to 
interpret the delay between them as direction of at least some flow 
of information.  

The application
to EEG and EMG data shows that the combination of both methods while
monitoring the SNR of all recording sites greatly improves the interpretation
of the results. In 
case of 
biological systems, there is a large variability (as revealed by surrogate
analysis) in the delay estimated by maximising coherence analysis. Though this
can partly be attributed to methodological problems, the major
reason for this variation may be the complex connections and the mode of
interaction between cortical and subcortical centers
\cite{williams,marsden1,marsden2}.  Application of time
delay analysis to complex (theoretical) networks may help to understand the
delay results of biological systems. This will be pursued in our future work. 
 
Even though there is a large variability in the magnitude of the delay,
direction of information flow is clearly revealed by the delay
analysis. However, the whole idea of using delay analysis to identify the
connection and interaction is valid only when there is a
\emph{non-zero} delay between two signals. In case of zero delay
between two signals, and to find out the direct and indirect
  coupling, one has to opt for methods like \emph{Extended   
  Granger Causality} \cite{gr} to identify the connections and 
interactions between the signals but this is beyond the scope of the current work.     
\label{sec9}

\section{Acknowledgement}

\label{sec10}

This work is supported by Deutsche Forschungsgemeinschaft (German Research
Council). We would like to thank Mrs. Lange for her assistance in recording 
high quality data.


\newpage
\begin{center}
Figure captions
\end{center}
Fig. 1 Results of delay, coherence and partial coherence analyses for model
system. In {\bf{b,c,f}} coherence between the time series
$X$ and $Y$ is displayed and is indicated by $X-Y$. 
In {\bf{d,g,h}} partial 
coherence between the time series $X$ and $Y$ accounting for the common
influence of the third time series $Z$ is displayed and is indicated by
$X-Y/Z$. Horizontal dotted lines in  {\bf{b,c,f}} and {\bf{d,g,h}} represent
the 99\% confidence limit for coherence and partial
coherence, respectively. In {\bf{a,e,i}} results of the delay analysis 
between the time series $X$ and $Y$ is displayed and is indicated by
$X-Y$. Here, as well as in the foregoing discussions, negative values of $\tau$
indicate that the time series of $X$ is 
shifted  
backwards in time to identify the delayed flow of information from $Y$ to $X$. In
{\bf{d}} partial coherence between $X_1(t)$ and $X_2(t)$ accounting for the
common influence of $X_3(t)$, is insignificant indicating that $X_1(t)$ and
$X_2(t)$ are communicating mainly through $X_3(t)$. {\bf{a}} There is a delay of 3 ms from $X_2(t)$
to $X_1(t)$. {\bf{e}} There is a delay of 2 ms from $X_1(t)$ to
$X_3(t)$. {\bf{i}} There is a delay of 5 ms from $X_2(t)$ to
$X_3(t)$. {\bf{j}} Connections and the direction of interactions used in
the 
model. {\bf{k}} Main connections  
identified by partial coherence analysis. However, though there is a
significant (partial) coherence in (g,h), a reduction compared to the
coherence in (c,f) indicates a weaker direct connection also between $X_1(t)$
and $X_2(t)$.  {\bf{l}} Connections and the
directions of interaction identified by delay
analysis.

Fig. 2 Results of delay, coherence and partial coherence analyses for 
ET1. Quantities plotted in {\bf{a-i,k,l}} have the same meaning as in
Fig. 1. In {\bf{b,c,f}} there is a significant coherence between the time
series 
$X$ and $Y$
(indicated by $X-Y$) at the tremor frequency of 4 Hz
which is shown is in {\bf{j}}. In {\bf{d}} partial coherence between F2 and C2 accounting for the
common influence of EMG, is insignificant (at the tremor of 4 Hz) indicating
that the connection between the SMA (F2) and MC (C2) is established mainly
through the EMG. In {\bf{g,h}} there is a significant partial cortico-muscular
when the influence of the other cortical area is removed. However the slight
reduction compared to the coherence in (c,f) 
indicates a weaker direct connection also
between F2 
and C2.  {\bf{a}} There is a delay of  7 ms from
C2 to F2 and 3 ms from F2 to C2. {\bf{e}} There is a delay of 19 ms from F2 to
EMG. {\bf{i}} There is a delay of 11 ms from C2 to EMG and 9 ms from EMG to
C2. For the sake of clarity, the abscissas in {\bf{a,e,i}} are plotted
in different scales.

Fig. 3 Results of delay, coherence and partial coherence analyses for 
ET2. Quantities plotted in {\bf{a-i,k,l}} have the same meaning as in
Fig. 1. In {\bf{b,c,f}} there is a significant coherence between the time
series 
$X$ and $Y$
(indicated by $X-Y$) at the tremor frequency of 5 Hz
which is shown in {\bf{j}}. In {\bf{d}} partial coherence between FCZ and C4 accounting for the
common influence of EMG, is insignificant (at the tremor of 5 Hz) indicating
that EMG is 
mainly influencing the SMA (FCZ) and MC (C4). In {\bf{g,h}} there is a
significant partial cortico-muscular coherence when the influence of the other
cortical area is removed, but it has reduced compared to the cortico-muscular
coherence (shown in c,f) indicating an indirect connection also between the two
cortical areas. {\bf{a}} There is a delay of  3 ms from
FCZ to C4. {\bf{d}} There is a delay of 25 ms from FCZ to
EMG. {\bf{i}} There is a delay of 20 ms from C4 to EMG. For the sake of
clarity, the abscissas in
{\bf{a,e,i}} are  plotted in different scales.

Fig. 4 Results of delay, coherence and partial coherence analyses for 
hs1. Quantities plotted in {\bf{a-i,k,l}} have the same meaning as in
Fig. 1. In {\bf{j}} there is a significant coherence at the tapping frequency
of 4 Hz between the EMG recorded
from both fingers. In {\bf{b,c,f}} there is a significant coherence
between the time series 
$X$ and $Y$ (indicated by $X-Y$) at the tapping (coherent)
frequency of 4 Hz which is shown in {\bf{j}}. Also, there is a significant
coherence in all the other frequency bands. In {\bf{d}} partial coherence between C3 and C4 accounting for the
common influence of CPZ, is insignificant (at the tapping frequency of 4 Hz as
well as in the other frequency bands) indicating
that the connection between C3 and C4 is established mainly through CPZ.
In {\bf{g,h}} there is a significant partial cortico-cortical coherence
between left/right hemisphere with the midline region
when the influence of the other hemisphere is removed. However, a
slight reduction in the partial coherence compared to the coherence shown in
(c,f) 
indicates that these two hemispheres are also (weakly) coupled.
{\bf{a}} There is a delay of  6 ms from
C4 to C3. {\bf{e}} There is delay of 2-3 ms from CPZ to
C3. {\bf{i}} There is a delay of 3-4 ms from C4 to
CPZ. For the sake of clarity, the abscissas in {\bf{a,e,i}} are plotted in different scale.

Fig. 5 Results of delay, coherence and partial coherence analyses for 
hs2. Quantities plotted in {\bf{a-i,k,l}} have the same meaning as in
Fig. 1. In {\bf{j}} there is a significant coherence at the tapping frequency
of 4 Hz between the EMG recorded
from both fingers. In {\bf{b,c}} there is a significant coherence between the
time series $X$ and $Y$ (indicated by $X-Y$) at the
tapping (coherent) 
frequency of 4 Hz (as well as at the double the tapping frequency) which is
shown is in {\bf{j}}.  In {\bf{f}} there is a significant
coherence at double the tapping frequency, 6-7 Hz. In {\bf{h}} partial
coherence between C3 and C4 accounting for the 
common influence of FCZ, is insignificant indicating
that the connection between C3 and C4 is established mainly through FCZ.
In {\bf{d,g}} there is significant partial coherence between left/right
hemisphere with the central region when the influence of the other hemisphere
is removed but the slight reduction (compared to the coherence shown in b,c)
indicates that these  hemispheres are also (weakly) coupled.  
However, though there is a
significant (partial) coherence in (d,g), a reduction compared to the
coherence (at double the tapping frequency) in (b,c) indicates a weaker direct connection also between C3
and C4. 
{\bf{a}} There is a delay of 6 ms from
C3 to FCZ and a delay of 15 ms from FCZ to C3. {\bf{e}} There is a delay of 16 ms from FCZ to
C4. {\bf{i}} There is a delay of 14 ms from C4 to
C3. For the sake of clarity, the abscissas in {\bf{a,e,i}} are plotted in
different scale.


\begin{thebibliography}{999}

\harvarditem[Albo et al.]{Albo et al.}{2004}{albo}
Albo Z, Prisco GVD, Chen Y, Rangarajan G, Truccolo W, Feng J, Vertes RP, Ding
M (2004) Is partial coherence a viable technique for identifying generators of
neural oscillations? Biol Cybern 90:318-326
\harvarditem[Amjad et al.]{Amjad et al.}{1997}{amj1}
Amjad AM, Halliday DM, Rosenberg JR, Conway BA (1997) An extended difference
of coherence test for comparing and combining several independent coherence
estimates:theory and application to the study of motor units and physiological
tremor. J Neurosci Methods 73:69-79
\harvarditem[Andres et al.]{Andres et al.}{1999}{andres}
Andres FG, Mima T, Schulman AE, Dichgans J, Hallett M, Gerloff C (1999)
Functional coupling of human cortical sensorimotor areas during bimanual skill
acquisition. Brain 122:855-870
\harvarditem[Baccala and Sameshima]{Baccala and Sameshima}{2001}{baccala}
Baccala LA, Sameshima K (2001) Partial directed coherence: a new concept in
neural structure determination. Biol Cybern 84:463-474
\harvarditem[Benabid et al.]{Benabid et al.}{1991}{benabid}
Benabid AL, Pollak P, Gervason C, Hoffmann D, Gao DM, Homme M, Perret JE
(1991) Long-term suppression of tremor by chronic stimulation of the ventral
intermediate thalamic nucleus. The Lancet 337:403-406
\harvarditem[Bernasconi and K\"onig]{Bernasconi and K\"onig}{1999}{berna}
Bernasconi C, K\"onig P (1999) On the directionality of cortical interactions
studied by structural analysis of electrophysiological
recordings. Biol Cybern 81:199-210  
\harvarditem[Brillinger]{Brillinger}{1981}{brillinger}
Brillinger DR (1981) Time series - Data analysis and theory, 2nd edn. Holden
Day, San Francisco
\harvarditem[Brinkman and Kuypers]{Brinkman and Kuypers}{1973}{brinkman}
Brinkman J, Kuypers HG (1973) Cerebral control of contralateral and
ipsilateral arm, hand and finger movements in the split-brain rhesus
monkey. Brain 96:653-674
\harvarditem[Bucher]{Bucher et al.}{1997}{bucher}
Bucher SF, Seelos KC, Dodel RC, Reiser M, Oertel WH (1997) Activation mapping
in essential tremor with functional magnetic resonance imaging. Ann Neurol
41:32-40 
\harvarditem[Carter]{Carter}{1987}{carter}
Carter GC (1987) Coherence and time delay estimation. IEEE 75:236-255
\harvarditem[Civardi et al.]{Civardi et al.}{2001}{civardi}
Civardi C, Cantello R, Asselman P, Rothwell JC (2001) Transcranial magnetic
stimulation can be used to test connections to primary motor areas from
frontal and medial cortex in humans. NeuroImage 14:1444-1453
\harvarditem[Chen et al.]{Chen et al.}{2004}{gr}  
Chen Y, Rangarajan G, Feng J, Ding M (2004) Analysing multiple nonlinear time series with extended Granger causality. Phys Lett A 324:26-35
\harvarditem[Dahlhaus]{Dahlhaus}{2000}{dahl1}
Dahlhaus R (2000) Graphical interaction models for multivariate time
series. Metrika 51:157-172
\harvarditem[Deuschl et al.]{Deuschl et al.}{1998}{deuschl}
Deuschl G, Bain P, Brin M (1998) Consensus statement of the movement disorder
society on tremor. Mov Disord 13:2-23
\harvarditem[De Gennaro et al.]{De Gennaro et al.}{2004}{gennaro1}
De Gennare L, Bertini M, Pauri F, Cristiani R, Curcio G, Ferrara M, Rossini PM
(2004) Callosal effects of transcranial magnetic stimulation (TMS):the
influence of gender and stimulus parameters. Neurosci Res 48:129-137
\harvarditem[De Gennaro et al.]{De Gennaro et al.}{2004}{gennaro2}
De Gennaro L, Ferrara M, Bertini M, Pauri F, Cristiani R, Curcio G, Romei V,
Fratello F, Rossini PM (2003) Reproducibility of callosal effects of
transcranial magnetic stimulation (TMS) with interhemispheric paired
pulses. Neurosci Res 46:219-227 
\harvarditem[Donchin et al.]{Donchin et al.}{2001}{donchin}
Donchin O, Gribova A, Steinberg O, Bergman H, Cardoso de Oliverira S, Vaadia E
(2001) Local field potentials related to bimanual movements in the primary and
supplementary motor cortices. Exp Brain Res 140:46-55
\harvarditem[Eliassen et al.]{Eliassen et al.}{2000}{eliassen}
Eliassen JC, Baynes K, Gazzaniga MS (2000) Anterior and posterior callosal
contributions to simultaneous bimanual movements of the hands and
fingers. Brain 123:2501-2511
\harvarditem[Findley and Koller]{Findley and Koller}{1987}{findley}
Findley LJ, Koller WC (1987) Essential tremor:a review. Neurology 37:1194-1197
\harvarditem[Gersch and Goddard]{Gersch and Goddard}{1970}{ger}
Gersch W, Goddard GV (1970) Epileptic focus location:spectral analysis
method. Science 169:701-702
\harvarditem[Govindan et al.]{Govindan et al.}{2005}{govindan}
Govindan RB, Raethjen J, Kopper F, Claussen JC, Deuschl G (2005) Estimation of
time delay by coherence analysis. Physica A 350:277-295
\harvarditem[Gribova et al.]{Gribova et al.}{2002}{gribova}
Gribova A, Donchin O, Bergman H, Vaadia E, Cardosa de Olivera S (2002) Timing
of bimanual movements in human and non-humans primates in relation to neuronal
activity in primary motor cortex and supplementary motor area. Exp Brain Res 146:322-335
\harvarditem[Hallett and Dubinsky]{Hallett and Dubinsky}{1993}{hallett}
Hallett M, Dubinsky RM (1993) Glucose metabolism in the brain of patients with
essential tremor. J Neurol Sci 114:45-58
\harvarditem[Halliday et al.]{Halliday et al.}{1995}{hal1}
Halliday DM, Rosenberg JR, Amjad AM, Breeze P, Conway BA, Farmer SF (1995) A frame work for the analysis of mixed time series/point
process data-theory and application to the study of physiological tremor,
single motor unit discharges and electromyograms. Prog Biophys Mol Biol
64:237-278
\harvarditem[Halliday et al.]{Halliday et al.}{1998}{hal2}
Halliday DM, Conway BA, Farmer SF, Rosenberg JR (1998) Using
electroencephalography to study functional coupling between cortical activity
and electromyograms during voluntary contractions in humans. Neurosci Lett
241:5-8 
\harvarditem[Hellwig et al.]{Hellwig et al.}{2001}{hellwig1}
Hellwig B, H\"au\ss ler S, Schelter B, Lauk M, Guschlbauer G, Timmer J,
L\"ucking CH (2001) Tremor-correlated cortical activity in essential
tremor. The Lancet 357:519-523
\harvarditem[Hjorth]{Hjorth}{1991}{hj1}
Hjorth B (1991) Principles for transformation of scalp EEG from potential
field into source distribution. J Clin Neurophysiol 8:391-396
\harvarditem[Hjorth]{Hjorth}{1975}{hj2}
Hjorth B (1975) An on-line transformation of EEG scalp potentials into
orthogonal source derivations. Electroencephalogr Clin Neurophysiol 39:526-530
\harvarditem[Honerkamp]{Honerkamp}{1994}{honerkamp}
Honerkamp J (1994) Stochastic dynamical systems. New York:VCH
\harvarditem[Hua et al.]{Hua et al.}{1998}{hua}
Hua SE, Lenz FA, Zirh TA, Dougherty PM (1998) Thalamic neuronal activity
correlated with essential tremor. J Neurosurg Psychiatry 64:273-276
\harvarditem[Immisch et al.]{Immisch et al.}{2001}{immisch}
Immisch I, Waldvogel D, Gelderen Pv, Hallett M (2001) The role of the medial
wall and its anatomical variations for bimanual antiphase and in-phase
movements. Neuroimage 14:674-684
\harvarditem[J\"ancke et al.]{J\"ancke et al.}{2000}{jancke}
J\"ancke L, Peters M, Himmelbach M, N\"osselt T, Shah J, Steinmetz H (2000)
fMRI study of bimanual coordination. Neuropsychologia 38:164-174
\harvarditem[Jenkins et al.]{Jenkins et al.}{1993}{jenkins}
Jenkins IH, Bain PG, Colebatch JG, Thompson PD, Findley LJ, Frackowiak RSJ,
Marsden CD, Brooks DJ (1993) A positron emission tomography study of essential
tremor:evidence for overactivity of cerebellar connections. Ann Neurol 34:82-90
\harvarditem[Kami\'{n}ski et al.]{Kami\'{n}ski et al.}{2001}{kaminski}
Kami\'{n}ski M, Ding M, Truccolo WA, Bressler SL (2001) Evaluating causal
relations in neural systems:granger causality, directed transfer function and
statistical assessment of significance. Biol Cybern 85:145-157
\harvarditem[Kantz and Schreiber]{Kantz and Schreiber}{1997}{kantz}
Kantz H, Schreiber T (1997) Nonlinear time series analysis. Cambridge
University Press.
\harvarditem[Kazennikov et al.]{Kazennikov et al.}{1999}{kazennikov}
Kazennikov O, Hyland B, Corboz M, Babalian A, Rouiller EM, Wiesendanger M
(1999) Neural activity of supplementary and primary motor areas in monkeys and
its relation to bimanual and unimanual movement sequences. Neurosci 89:661-674
\harvarditem[Klem et al.]{Klem et al.}{1999}{klem}
Klem GH, Jasper HO, Elgerl HH (1999) The ten-twenty electrode system of the
International Federation, in: Deuschl G, Eisen A (ed) Recommendations for the
practice of clinical Neurophysiology: Guidelines of the International
Federation of Clinical Neurophysiology. Electroencephalogr Clin Neurophys
Suppl 52:3- 
\harvarditem[Kocsis et al.]{Kocsis et al.}{1999}{kocsis}
Kocsis B, Bragin A, Buzaki G (1999) Interdependence of multiple theta
generators in the hippocampus: a partial coherence analysis. J Neurosci
19:6200-6212
\harvarditem[Kubota et al.]{Kubota et al.}{2003}{kubota}
Kubota D, Colgin LL, Casale M, Brucher FA, Lynch G (2003) Endogenous waves in
hippocampal slices. J Neurophysiol 89:81-89
\harvarditem[Kus et al.]{Kus et al.}{2004}{kus}
Ku\'{s} R, Kami\'{n}ski M, Blinowska K.J. (2004) Determination of EEG activity propagation:Pair-Wise versus multichannel estimate.
IEEE Tran Biomed Eng 51:1501-1510
\harvarditem[Lamarre]{Lamare}{1984}{lamarre}
Lamarre Y (1984) Animal models of physiological, essential and
Parkinsonian-like tremors. Findley LJ and Capildeo R (ed) Movement
disorders:tremor. Macmillan Press, London:183
\harvarditem[Lang et al]{Lang et al.}{1990}{lang}
Lang W, Obrig H, Lindinger G, Cheyne D, Deecke L (1990) Supplementary motor
area activation while tapping bimanually different rhythms in musicians. Exp
Brain Res 79:504-514
\harvarditem[Liang et al.]{Liang et al.}{1998}{liang}
Liang H, Ding M, Nakamura R, Bressler SL (2000) Causal influence in primate
cerebral cortex during visual pattern discrimination. Neuroreport 11:2875-2880
\harvarditem[Llin$\rm{\acute{a}}$s and Volkind]{Llin$\rm{\acute{a}}$s and
Volkind}{1973}{llinas}  
Llin$\rm{\acute{a}}$s R, Volkind RA (1973) The olivocerebellar system:
functional properties as revealed by harmaline-induced tremor. Exp Brain Res
18:69-87
\harvarditem[Lopes et al.]{Lopes et al.}{1980}{lopes}
Lopes da Silva FH, Vos JE, Mooibroek J, Van Rotterdam A (1980) Relative
contributions of intracortical and thalamocortical processes in the generation
of alpha rhythms, revealed by partial coherence analysis. Electroencephalogr
Clin Neurophysiol 50:449-456 
\harvarditem[Marsden et al.]{Marsden et al.}{2000}{marsden1}
Marsden JF, Ashby P, Limousin-Dowsey P, Rothwell JC, Brown P (2000) Coherence
between cerebellar thalamus, cortex and muscle in man. Brain 123:1459-1470
\harvarditem[Marsden et al.]{Marsden et al.}{2001}{marsden2}
Marsden JF, Limousin-Dowsey P, Ashby P, Pollak P, Brown P (2001) Subthalamic
nucleus, sensorimotor cortex and muscle interrelationships in Parkinson's
disease. Brain 124:378-388
\harvarditem[Meyer et al.]{Meyer et al.}{1995}{meyer}
Meyer BU, Roricht S, Grafin von Einsiedel H, Kruggel F, Weindl A (1995)
Inhibitory and excitatory interhemispheric transfers between motor cortical
areas in normal humans and patients with abnormalities of the corpus
callosum. Brain 118:429-440
\harvarditem[Mima et al.]{Mima et al.}{2000a}{mima1}
Mima T, Matsuoka T, Hallett M (2000) Functional coupling of human right and
left cortical motor areas demonstrated with partial coherence
analysis. Neurosci Lett 287:93-96 
\harvarditem[Mima and Hallett]{Mima and Hallett}{1999a}{mima2}
Mima T, Hallett M (1999a) Cortico-muscular coherence: a review.J Clin
Neurophysiol 16:501-511
\harvarditem[Mima and Hallett]{Mima and Hallett}{1999b}{mima3}
Mima T, Hallett M (1999b) Electroencephalographic analysis of cortico-muscular
coherence:reference effect, volume conduction and generator mechanism. Clin
Neurophysiol 110:1892-1899
\harvarditem[Mima et al.]{Mima et al.}{2000b}{mima4}
Mima T, Steger J, Schulman AF, Gerloff C, Hallett M (2000)
Electroencephalographic measurement of motor cortex control of muscle activity
in humans. Clin Neurophysiol 111:326-337
\harvarditem[M\"uller et al.]{M\"uller}{2003}{timmer}
M\"uller T, Lauk M, Reinhard M, Hetzel A, L\"ucking CH, Timmer J (2003)
Estimation of time-delays in biological systems. Annals Biomed Eng
31:1423-1439 
\harvarditem[Pollok et al.]{Pollok et al.}{2004}{pollok}
Pollok B, Gross J, Dirks M, Timmermann L, Schnitzler A (2003) The cerebral
oscillatory network of voluntary tremor. J Physiol 554:871-878
\harvarditem[Raethjen et al.]{Raethjen et al.}{2005}{raethjen2}
Raethjen J, Govindan RB, Kopper F, Deuschl G (2005) Cortical involvement in
the generation of essential tremor. (submitted)
\harvarditem[Raethjen et al.]{Raethjen et al.}{2004}{raethjen1}
Raethjen J, Lindemann M, Morsnowski A, Dumpelmann M, Wenzelburger R, Stolze H,
Fietzek U, Pfister G, Elger CE, Timmer J, Deuschl G (2004) Is the rhythm of
physiological tremor involved in cortico-cortical interactions? Mov
Disord 19:458-465
\harvarditem[Rosenberg et al.]{Rosenberg et al.}{1998}{ros1}
Rosenberg JR, Halliday DM, Breeze P, Conway BA (1998) Identification of
patterns of neuronal connectivity-partial spectra, partial coherence, and
neuronal interactions. J Neurosci Methods 83:57-72
\harvarditem[Rothwell et al.]{Rothwell et al.}{1991}{rothwell}
Rothwell JC, Thompson PD, Day BL, Boyd S, Marsden CD (1991) Stimulation of the
human motor cortex through the scalp. Exp Physiol 76:159-200
\harvarditem[Rybski et al.]{Rybski et al.}{2003}{diego}
Rybski D, Havlin S, Bunde A (2003) Phase synchronisation in temperature and
precipitation records. Physica A 320:601-610
\harvarditem[Schmierer et al.]{Schmierer et al.}{2000}{schmierer}
Schmierer K, Niehaus L, R\"oricht, Meyer BU (2000) Conduction deficits of
callosal fibres in early multiple sclerosis. J Neurosurg Psychiatry 68:633-638
\harvarditem[Schuurman et al.]{Schuurman et al.}{2000}{schuurman}
Schuurman PR, Bosch DA, Bossuyt PMM, Bonsel GJ, van Someren EJW, de Bie RMS,
Merkus MP, Speelman JD (2000) A comparison of continuous thalamic stimulation
and thalamotomy for suppression of severe tremor. N Engl J Med 342:461-468
\harvarditem[Sherman et al.]{Sherman et al.}{1997}{sherman}
Sherman DL, Tsai YC, Rossell LA, Mirski MA, Thakor NV (1997) Spectral analysis
of a thalamus-to-cortex seizure pathway. IEEE Trans Biomed Eng 44:657-664
\harvarditem[Stephan et al.]{Stephan et al.}{1999}{stephan}
Stephan KM, Binkofski F, Halsband U, Dohle C, Wunderlich G, Schnitzler A, Tass
P, Posse S, Herzog H, Sturm V, Zilles K, Seitz RJ, Freund HJ (1999) The role
of ventral medial wall motor areas in bimanual co-ordination. A combined
lesion and activation study. Brain 122:351-368
\harvarditem[Theiler et al.]{Theiler et al.}{1992}{theiler}
Theiler J, Longtin A, Galdrikan B, Farmer JD (1992) Testing for nonlinearity
in time series: The method of surrogate data. Physica D 58:77-94
\harvarditem[Timmer et al.]{Timmer et al.}{1998a}{timmer1}
Timmer J, Lauk M, Pfleger W, Deuschl G (1998a) Cross-spectral analysis of
physiological tremor and muscle activity. I: Theory and application to
unsynchronised EMG. Biol Cybern 78:349-357
\harvarditem[Timmer et al.]{Timmer et al.}{1998b}{timmer2}
Timmer J, Lauk M, Pfleger W, Deuschl G (1998b)  Cross-spectral analysis of
physiological tremor and muscle activity. II: Theory and application to
synchronised EMG. Biol Cybern 78:359-368
\harvarditem[Timmermann et al.]{Timmermann et al.}{2003}{timmermann}
Timmermann L, Gross J, Dirks M, Volkmann J, Freund HJ, Schnitzler A (2003) The
cerebral oscillatory network of Parkinsonian tremor. Brain 126:199-212
\harvarditem[Turbes and Schneider]{Turbes and Schneider}{1989}{turbes}
Turbes CC, Schneider GT (1989) Directionality of neural signals in central
nervous system neural networks. Biomded Sci Instrum 25:1-5
\harvarditem[Urbano et al.]{Urbano et al.}{1998}{urbano}
Urbano A, Babiloni C, Onorati P, Carducci F, Ambrosini A, Fattorini L,
Babiloni F (1998) Responses of human primary sensorimotor and supplementary
motor areas to internally triggered unilateral and simultaneous bilateral one
digit movements. A high resolution EEG study. Eur J Neurosci 10:765-770
\harvarditem[Vertes et al.]{Vertes et al.}{2001}{vertes}
Vertes RP, Albo Z, Viana Di Prisco G (2001) Theta rhythmically firing neurons
in the anterior thalamus: implications for 
the mnemonic function of Papey's circuit. Neuroscience 104:619-625
\harvarditem[Williams et al.]{Williams et al.}{2002}{williams}
Williams D, Tijssen M, Bruggen Gv, Bosch A, Insola A, Lazzaro VD, Mazzone P,
Oliviero A, Quartarone A, Speelman H, Brown P (2002) Dopamine-dependent
changes in the functional connectivity between basal ganglia and cerebral
cortex in humans. Brain 125:1558-1569

%








\end{thebibliography}
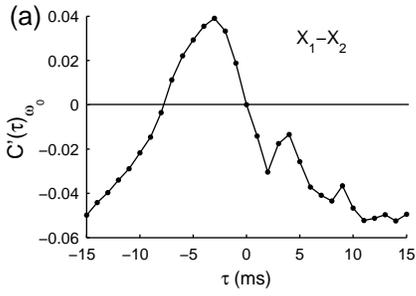
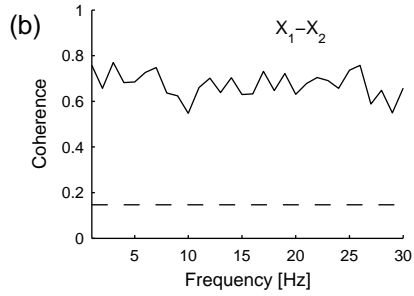
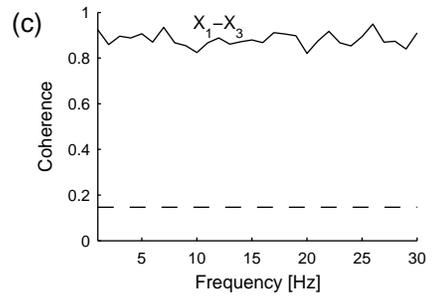
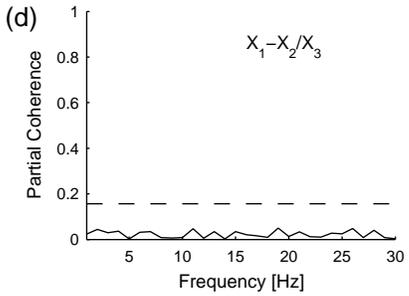
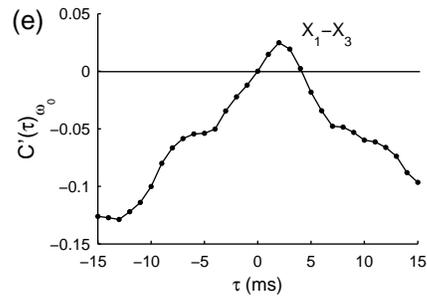
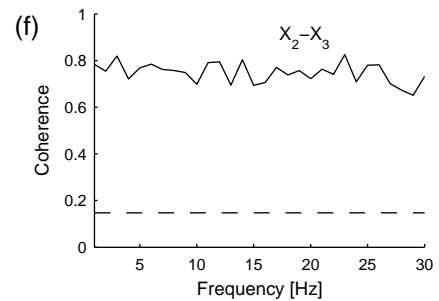
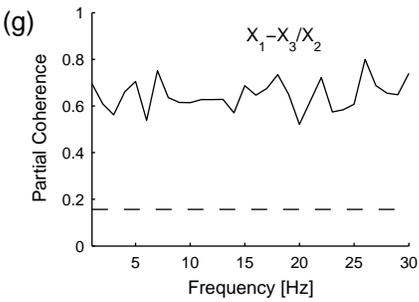
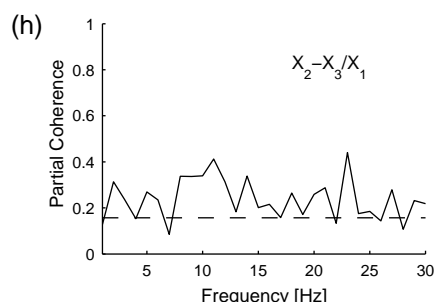
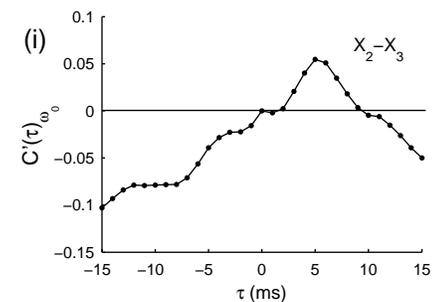
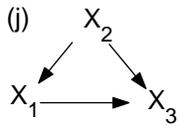
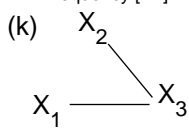
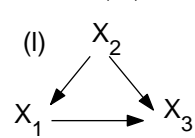
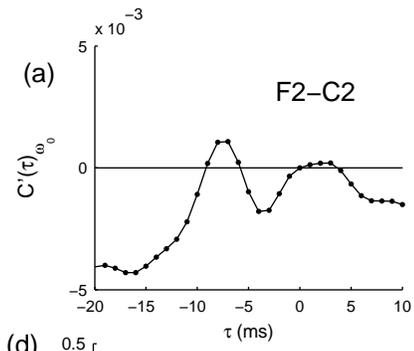
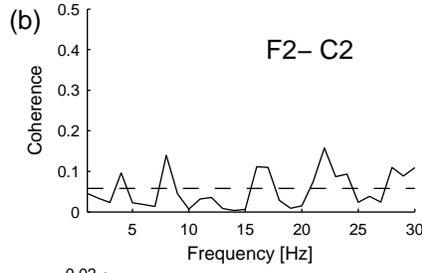
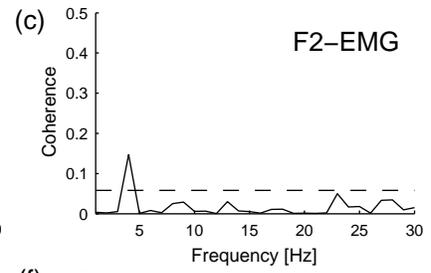
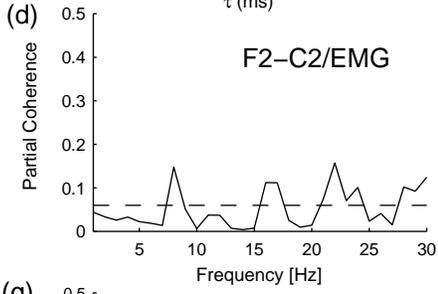
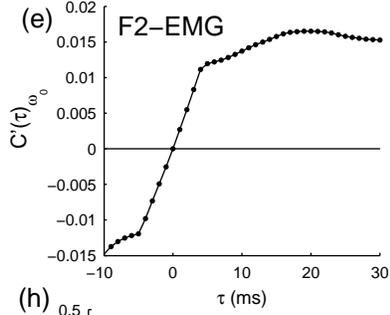
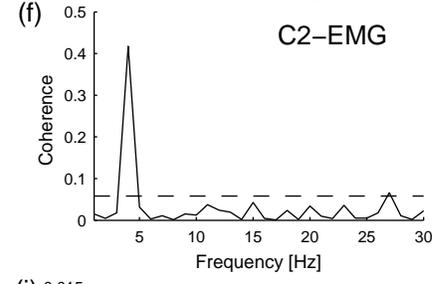
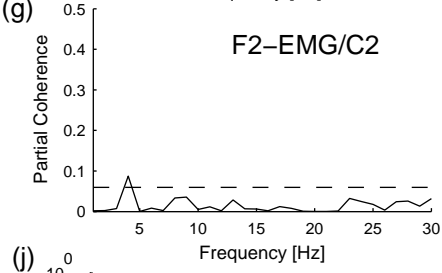
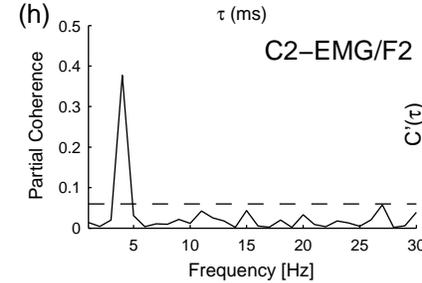
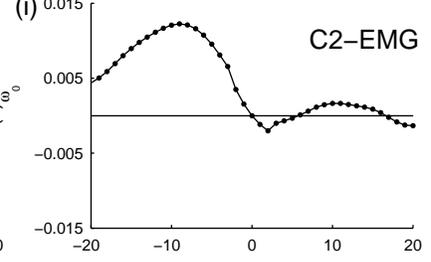
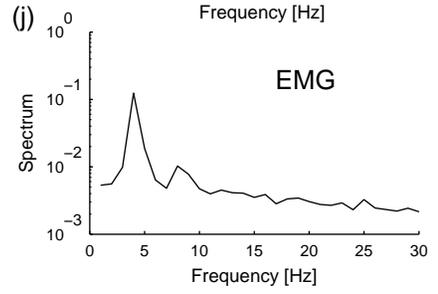
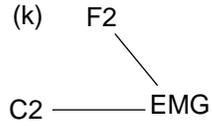
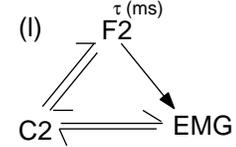
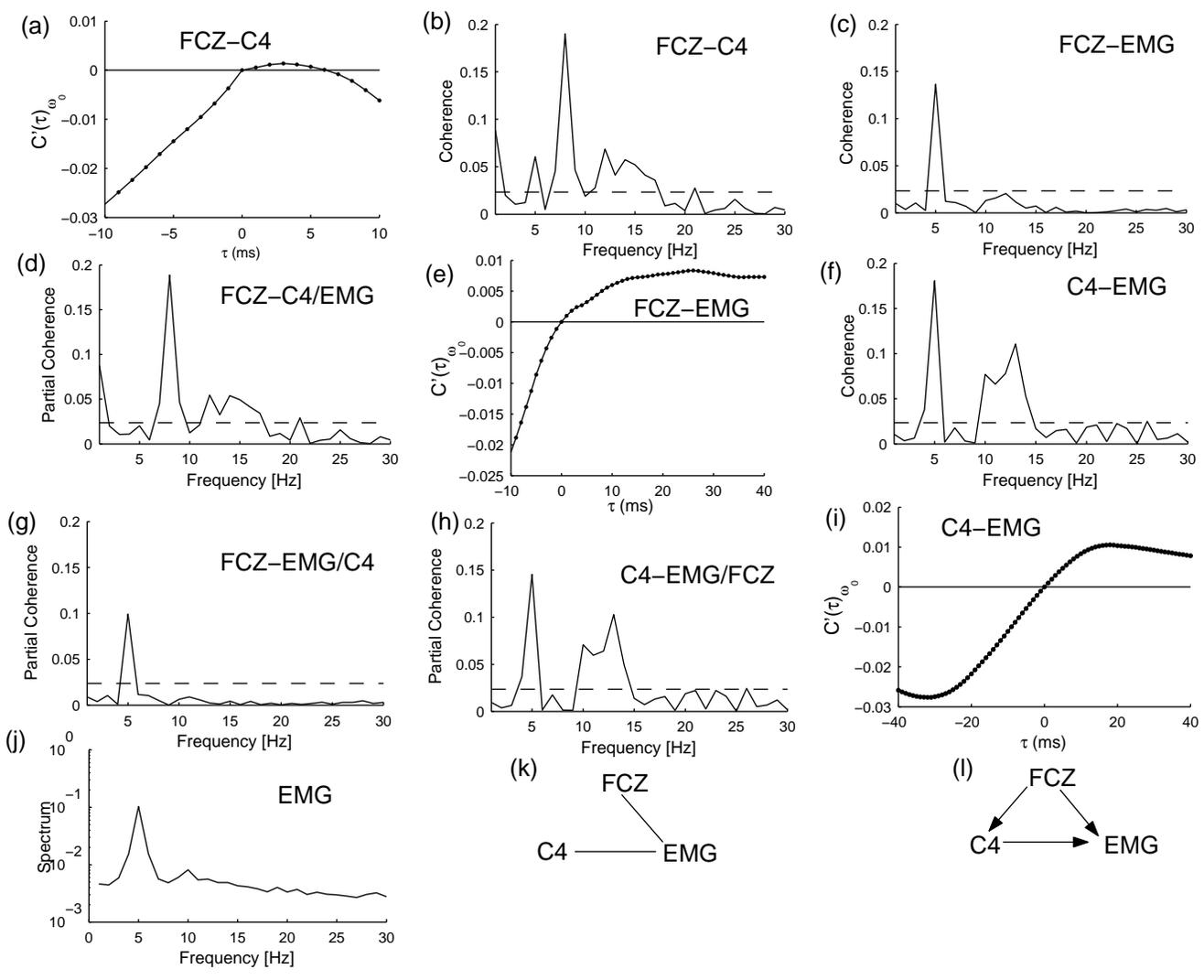
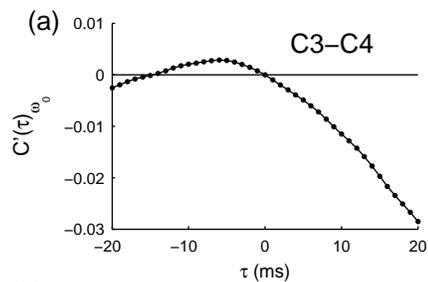
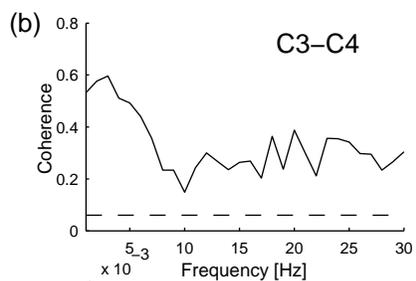
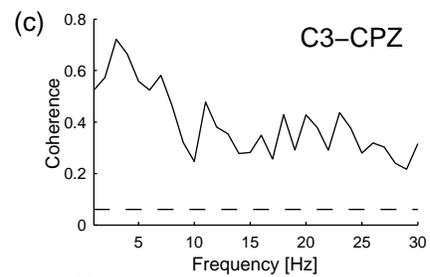
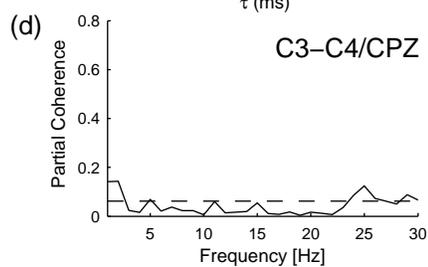
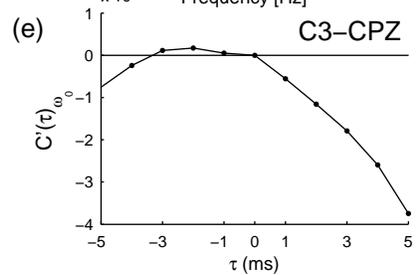
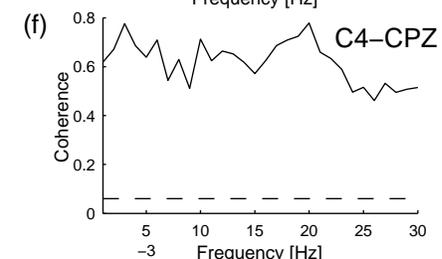
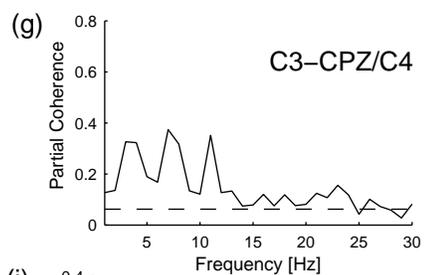
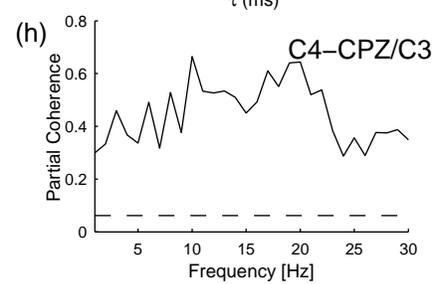
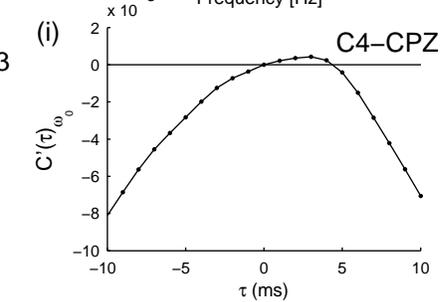
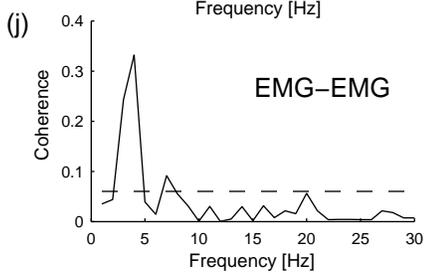
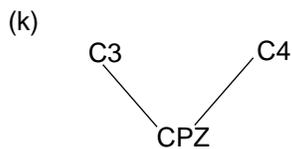
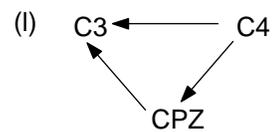
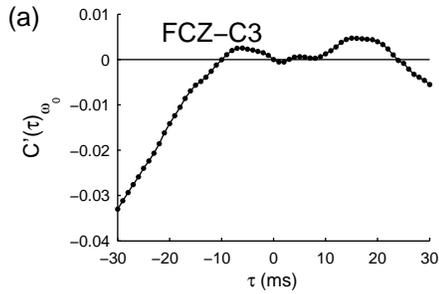
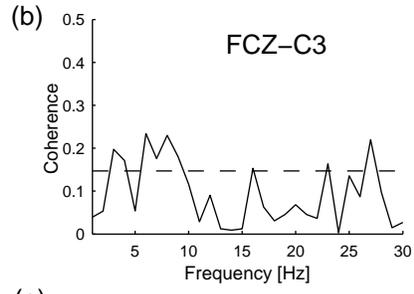
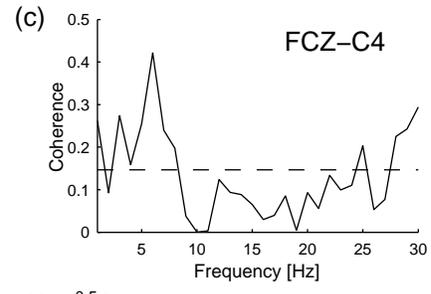
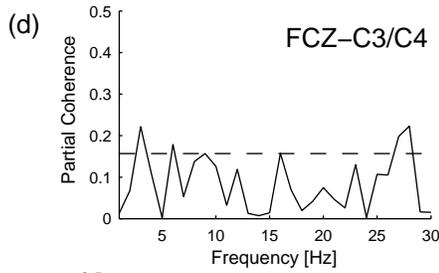
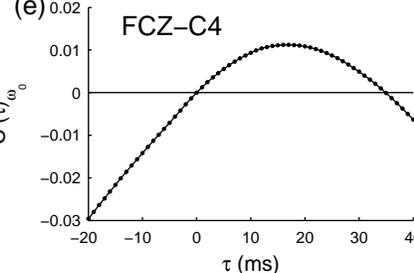
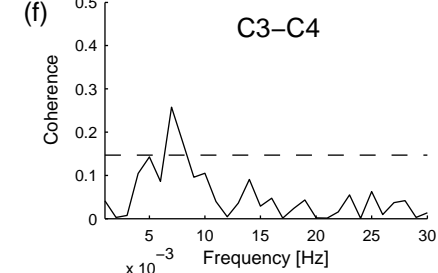
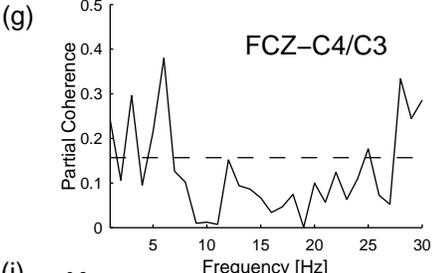
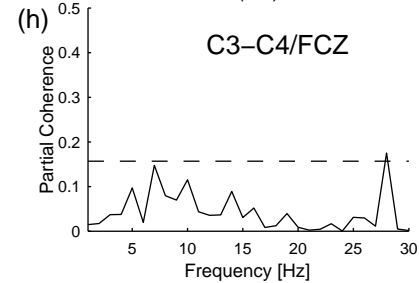
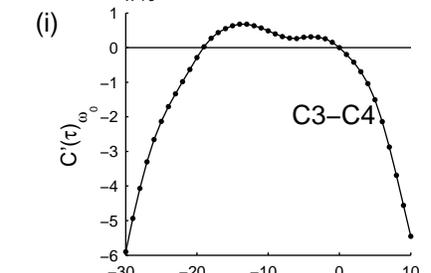
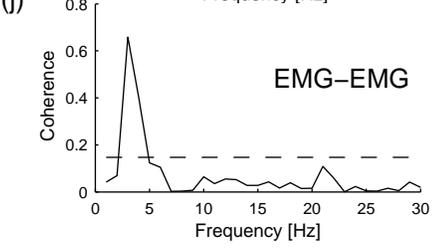
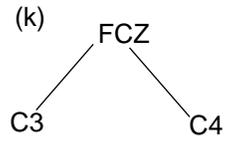
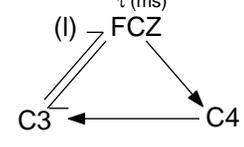
\end{document}